\documentclass{epl}
\usepackage{euscript}

\newcommand{\B}{\EuScript{B}}
\newcommand{\A}{\EuScript{A}}
\newcommand{\SpinSet}{\{\uparrow,\downarrow\}}
\newcommand{\sgn}{\mbox{sgn }}
\newcommand{\vol}{\mbox{Vol }}
\newcommand{\tr}{\mbox{tr }}

\title{A Proof of Luttinger's Theorem}% Force line breaks with \\

\author{A. Praz\inst{1,2} \and J. Feldman\inst{3} \and H. Kn\"orrer\inst{1}
\and E. Trubowitz\inst{1}}
\institute{
  \inst{1} Department of Mathematics, Eidgen\"ossische Technische Hochschule Z\"urich, 
           Z\"urich, Switzerland\\
  \inst{2} Paul Scherrer Institut, Villigen, Switzerland\\
  \inst{3} Department of Mathematics, University of British Columbia, Vancouver, Canada
}

\begin{document}

\maketitle

\begin{abstract}

 A rigorous and simple perturbative proof of Luttinger's theorem is sketched for Fermi liquids in two and three dimensions. It is proved that in the finite volume, the quasi-particle density is independent of the interaction strength. The thermodynamic limit is then controlled to all orders in perturbation
theory.

\end{abstract}

For more than the past 40 years Fermi liquid theory has played a major role in the description of condensed matter systems at low temperature. One of the most striking characteristics of a Fermi liquid is probably the presence of a Fermi surface, revealed by a discontinuity in the occupation number as function of the quasi-momentum at zero temperature. Luttinger found in his seminal paper~\cite{L} that the volume enclosed in the Fermi surface is independent of the interaction strength, if the density is held fixed. Assuming the existence of a formal expansion in powers of the interaction strength, Luttinger proved that the higher order contributions to the volume vanish order by order in the perturbation expansion.

From a mathematical point of view, the proof given by Luttinger is unsatisfactory. The manipulations of conditionally convergent integrals that lead to Luttinger's result have to be controlled rigorously. One could introduce an infrared cut-off, in which case Luttinger's argument is mathematically valid, and carefully remove the cut-off in a second step, using renormalization group ideas.

Recently, an alternative topological proof of Luttinger's theorem was given by Oshikawa~\cite{O}. On a finite torus, he considered the change in the total momentum in a particular direction under the adiabatic insertion of a unit quantum flux $\Phi_0=hc/e$ through the hole of the torus. Expressing the change in the total momentum in terms of the volume of the Fermi sea, he showed that the volume of the Fermi sea is given (up to multiplication by a constant) by the particle density plus an integer, corresponding to the number of completely filled bands. As Luttinger did, Oshikawa assumed the existence of the thermodynamic limit, and completed the proof of Luttinger's theorem.

Instead of following the argument of Luttinger, we propose here a simpler proof of Luttinger's theorem. We first observe that the number operator commutes with the Hamiltonian. Thus, the ground state of the system in a finite spatial volume is an eigenstate of the number operator, whose eigenvalues are natural numbers. It follows that at zero temperature, if the ground state is non-degenerate, the expectation value of the number operator is an integer.
On the other hand, we show that for generic dispersion relations (we consider a single band system, as for example a Hubbard system far from half-filling), the expectation value at zero temperature of physical observables, like for instance the number operator, are analytic functions of the interaction strength in the finite volume. Obviously, continuous variation of the interaction strength cannot change the zero temperature expectation value of the number operator. We deduce that, in the finite volume, the particle density is given by a power series in the coupling constant, in which only the zero order Taylor coefficient is non-vanishing. If the thermodynamic limit of the expectation value of any physical observable is carefully controlled, the above statement remains true in the infinite volume.

In order to avoid unphysical infrared divergences in the thermodynamic limit, we need to renormalize the single particle dispersion relation. In our renormalization scheme, the ``interacting'' dispersion relation, defining the physical Fermi surface, appears in the propagator. We introduce a counter-term that holds the Fermi surface fixed, and obtain a renormalization map between the interacting dispersion relation, and the bare dispersion relation, determined by the single-particle Schr\"odinger equation. We now have all the ingredients needed to prove Luttinger's theorem, which is obtained by fixing the chemical potential to its physical value, both in the free fermion approximation, and in the interacting system. Comparison of the two results for the volume of the Fermi sea yields Luttinger's theorem.

In the present work, the thermodynamic limit is controlled order by order in powers of the interaction strength. In that sense, our result, as well as the proof of Luttinger~\cite{L} or Oshikawa~\cite{O} are perturbative, in contrast to the recent proof~\cite{FKT} of the existence of a 2D Fermi liquid at zero temperature, where the convergence of the power series in the interaction strength is established. Details of the proof of the existence of the thermodynamic limit in the sense of formal power series will be presented in a forthcoming paper by Praz~\cite{P}.

A necessary assumption for the validity of Luttinger's theorem seems to be gauge invariance, or equivalently charge conservation. The argument of Luttinger can be related to Ward identities. Oshikawa explicitly makes use of gauge invariance and we require particle number conservation.

We consider many fermions on the finite lattice of size $L^d$ with periodic boundary conditions, where the size of the unit cell is one. The system will be described by the many body Hamiltonian $H=H_0+\lambda V$, where the single-particle Hamiltonian is diagonal in momentum space, 
\begin{equation}
H_0=L^{-d}\sum_{{\bf k},\sigma}\varepsilon({\bf k})c_{{\bf k}\sigma}^+c_{{\bf k}\sigma}.\label{H0}
\end{equation}
Here ${\bf k}$ is in the dual lattice, and $\sigma\in \SpinSet$ is the spin. $\varepsilon({\bf k})$ is the one-particle, single-band dispersion relation. Strictly speaking, the dispersion relation should depend on the size of the system. In order to lighten the notations, we work directly with the dispersion relation $\varepsilon({\bf k})$, defined on the first Brillouin zone, neglecting unimportant terms of order $1/L$. Precise assumptions on the properties of $\varepsilon({\bf k})$ are given below. The creation and annihilation operators $c^+_{{\bf k}\sigma}$ and $c_{{\bf k}\sigma}$ satisfy the fermionic anticommutation relations $\{c^+_{{\bf k}\sigma},c_{{\bf p}\tau}\}=(2\pi L)^d \delta_{\bf kp}\delta_{\sigma\tau}$. The pair interaction between fermions is given by 
\begin{equation}
V=L^{-3d}\mathop{\sum_{{\bf k}_1,\dots,{\bf k}_4}}_{\sigma,\tau}\delta_{{\bf k}_1+{\bf k}_3,{\bf k}_2+{\bf k}_4}
v_{\sigma\tau}({\bf k}_1-{\bf k}_3)
c_{{\bf k}_1\sigma}^+c_{{\bf k}_2\tau}^+c_{{\bf k}_3\sigma}c_{{\bf k}_4\tau}
\end{equation}
where we shall assume that the potential $v({\bf k})$ is local, i.e., is a smooth function of ${\bf k}$. Our proof can trivially be extended to potentials depending smoothly on the frequency. 

The thermal expectation value at zero temperature of any polynomial in creation and annihilation operators $\A$ is 
\begin{equation}
\langle \A\rangle_{L,\lambda}=\frac{\tr P_{\lambda,\mu}^{(L)}\A }{\tr P_{\lambda,\mu}^{(L)}},
\end{equation}
where $P_{\lambda,\mu}^{(L)}$ is the projection on the ground state of $H_0-\mu N+\lambda V$ 
where the number operator $N=L^{-d}\sum_{{\bf k},\sigma}c_{{\bf k}\sigma}^+c_{{\bf k}\sigma}$, and $\mu$ is the chemical potential. In the thermodynamic limit $L\rightarrow \infty$, the naive power expansion of $\langle \A\rangle_{L,\lambda}$ in $\lambda$ does not exist. Precisely, the coefficients of $\lambda^n$ in the power expansion diverge as $L\rightarrow\infty$ for $n\ge 3$. The renormalization procedure cures these (unphysical) infrared divergences order by order in $\lambda$. 

In the present work, we are concerned with the perturbative theory only. We will not be interested in the convergence of the whole perturbative series, which is related to (the absence of) instability in the ground state. We thus consider only formal power series. $f(\lambda)=\sum_{n\ge 0}f_n\lambda^n$ is a well-defined formal power series in $\lambda$ if for all $n$, $|f_n|<\infty$. 

Infrared divergences appear because the bare propagator used in the expansion has a pole on the bare Fermi surface, while the physical pole lies on the interacting Fermi surface. We propose a renormalization scheme which allows a diagrammatic expansion around the interacting Fermi surface. Formally, we split the bare dispersion relation in a renormalized dispersion relation $E({\bf k};\mu,\lambda)$, and a counter-term $K({\bf k};\lambda)$ such that $\varepsilon({\bf k})-\mu = E({\bf k},\mu,\lambda)+K({\bf k};\lambda)$. The counter-term is fixed by the renormalization condition, which guarantees that the interacting Fermi surface is determined by the zero set of $E({\bf k},\mu,\lambda)$. Unfortunately, the renormalized dispersion relation cannot be determined directly. We thus begin with a generic dispersion relation $e({\bf k},\mu)$, and consider the model defined by the Hamiltonian
\begin{equation}
H^{(L)}=L^{-d}\sum_{{\bf k},\sigma}e^{(L)}({\bf k})c_{{\bf k}\sigma}^+c_{{\bf k}\sigma}+\lambda V + \EuScript{K}^{(L)},\label{HamiltonL}
\end{equation}
where
\begin{equation}
e^{(L)}({\bf k})=\left\{
\begin{array}{cl}
e({\bf k}), &\mbox{if $|e({\bf k})|\ge 4\pi / L$,}\\
\sgn(e({\bf k}))\frac{4\pi}{L}, &\mbox{if $|e({\bf k})|< 4\pi / L$},\\
4\pi/L, &\mbox{if $e({\bf k})=0$}
\end{array}\right.\label{RegDispRel}
\end{equation}
is a regularized dispersion relation (the $\mu$ dependency has been dropped for simplicity), and $\EuScript{K}^{(L)}=L^{-d}\sum_{{\bf k},\sigma}K^{(L)}({\bf k};e(\,\cdot\, ,\mu),\lambda)c_{{\bf k}\sigma}^+c_{{\bf k}\sigma}$ is the counter-term, to be determined by the (discretized) renormalization condition. We then control the thermodynamic limit of the expectation value of an operator $\A$, defined by 
\begin{equation}
\Xi^{(L)}(\A;e,\lambda)=\frac{\tr \A\Pi_{\lambda,e}^{(L)}}{\tr \Pi_{\lambda,e}^{(L)}}\label{XiExpectationValue}
\end{equation}
where $\Pi_{\lambda,e}^{(L)}$ is the projection onto the state of of minimum energy of the Hamiltonian~(\ref{HamiltonL}). In the infinite volume, the occupation number of this model has a jump on the surface defined by
\begin{equation}
\Sigma_{e,\mu}=\{{\bf k}:\, e({\bf k},\mu)=0 \},\label{FermiSurface} 
\end{equation}
which is identified as the interacting Fermi surface. If
\begin{equation}
\varepsilon({\bf k})-\mu =  e({\bf k},\mu)+K({\bf k};\lambda,e(\,\cdot\, ,\mu)),\label{ImplicitEquation}
\end{equation} 
we can identify the physical expectation value $\langle \A \rangle$ in the infinite volume with 
$\displaystyle\lim_{L\rightarrow\infty}\Xi^{(L)}(\A;e,\lambda)$. The solution of the implicit equation~(\ref{ImplicitEquation}) is obtained inverting the renormalization map $R_\lambda: e\mapsto e+K(\,\cdot\, ;e,\lambda)$, and corresponds to the sought-after interacting dispersion relation 
\begin{equation}
E(\,\cdot\, ,\lambda,\mu)=R^{-1}_\lambda(\varepsilon-\mu).\label{EdispersionRelation}
\end{equation} 
The expectation value $\langle \A\rangle$ can finally be expressed as
\begin{equation}
\langle \A \rangle_\lambda = \lim_{L\rightarrow\infty}
\left.\Xi^{(L)}(\A;E(\,\cdot\, ,\lambda,\mu),\lambda')\right|_{\lambda'=\lambda},
\end{equation}
defined as a (formal) power series in $\lambda'$. The power series expansion  of $\langle \A \rangle_\lambda$ in the coupling constant $\lambda$ does not exist, since the power series expansion of $\Xi^{(L)}(\A,e,\lambda')$ in the dispersion relation $e$ does not exist. For $\lambda=0$, one recovers the expectation value of a system of free fermions. 

In the traditional Wilsonian approach to the renormalization group, the high energy degrees of freedom are successively integrated out, modifying at each renormalization step the dispersion relation and the interaction. The bare Hamiltonian flows toward an effective Hamiltonian, which should describe the low energy physics. Our construction begins directly with the interacting dispersion relation $e$, with the advantage of using the same propagator at each energy scale. The price to pay is the non-trivial inversion of the renormalization map. It is worth noting that the Hamiltonian~(\ref{HamiltonL}) is not an effective Hamiltonian, but the starting point of the perturbation expansion. 

We now turn to the rigorous formulation of our results. Let $\B=\{e:({\bf p},\mu)\mapsto e({\bf p},\mu) \}$ be the space of all possible dispersion relations satisfying the following assumptions: (i) There is an open subset $U$ of the first Brillouin zone, such that for all chemical potentials $\mu\in I$, where $I$ is an interval of real numbers, $e(\,\cdot\, ,\mu)$ is smooth on $U$. (ii) For all $\mu\in I$, the surface defined by~(\ref{FermiSurface}) is contained in $U$, and $\nabla e({\bf k})\neq 0$ for all ${\bf k}\in \Sigma_{e,\mu}$ and $\mu\in I$. (iii) For all $\mu \in I$, the surface $\Sigma_{e,\mu}$ is strictly convex. We require that $\varepsilon({\bf k})-\mu$ satisfies the assumptions (i)-(iii).

The first and second assumptions ensure the regularity and smoothness of the Fermi surface; in particular, van Hove singularities are excluded by (ii). The third assumption is used to control the value of Feynman graphs with two or more loops. The dispersion relations in $\B$ will correspond to the renormalized (or interacting) dispersion relations. Assumptions (i)-(iii) are hypotheses used in the proof of the following theorem on the existence of the thermodynamic limit:

{\bf Theorem 1:}\label{Thm1}
 Let $e$ be a dispersion relation in $\B$, and $v_{\sigma\tau}$ be a local interaction. Then there is a neighborhood $\EuScript{O}$ of $e$ in $\B$, and a renormalization map 
\begin{equation}
\begin{array}{clcl}
R^{(L)}_\lambda:&\EuScript{O}&\rightarrow &\EuScript{O}\\
                & e          &\mapsto     &e + K^{(L)}(\,\cdot\, ;e,\lambda)
\end{array}
\end{equation}
such that the following holds. The expectation value of an operator $\A$, defined by~(\ref{XiExpectationValue}) is analytic in $\lambda$, with a radius of analyticity that may shrink to zero as $L\rightarrow\infty$. For $L\rightarrow \infty$, the renormalization map tends to an invertible map $R_\lambda: e\mapsto R_\lambda e$, defined as a formal power series in $\lambda$, such that the limit
\begin{equation}
\Xi(\A;e,\lambda)=\lim_{L\rightarrow\infty}\Xi^{(L)}(\A;e,\lambda)
\end{equation}
is well-defined in the sense of formal power series, i.e., for $\Xi^{(L)}(\A;e,\lambda)=\sum_{n\ge 0}\Xi_n^{(L)}(\A;e)\lambda^n$, 
\begin{equation}
\Xi_n^{(L)}(\A;e)\stackrel{L\rightarrow \infty}{\rightarrow}\Xi_n(\A;e)
\end{equation}
for all $n$. In particular, the occupation number, defined as
\begin{equation}
\eta_\sigma({\bf k};e,\lambda)=\lim_{L\rightarrow\infty} \Xi^{(L)}(c^+_{{\bf k}\sigma}c_{{\bf k}\sigma};e,\lambda)
\end{equation}
has a discontinuity as function of ${\bf k}$ on the Fermi surface $\Sigma_{e\mu}$. The counter-term is determined by the condition that the self-energy vanishes order by order on the surface $\Sigma_{e\mu}$.

A proof of this theorem will be given in a later paper~\cite{P}. 
The theorem~1 is a generalization for finite volumes of the main results of~\cite{FST}-\cite{FST4}.

The proof of the analyticity of the expectation values in the finite volume would be trivial without the counter-term. One has to show that the discretized renormalization condition has a solution, which is analytic in $\lambda$. Renormalization group ideas are used in order to control the value of Feynman graphs in the limit $L\rightarrow \infty$. The vicinity of the Fermi surface~(\ref{FermiSurface}) is split into shells fitting into each other, the width of the shells getting smaller near the Fermi surface. The contributions of Feynman graphs at fixed scales are  bounded by power counting, and careful summation over the scales shows the convergence of the Green's functions. When the width of the shell is of the order $1/L$, the power counting becomes worst, and scale decomposition proves to be useless. The regularized dispersion relation~(\ref{RegDispRel}) provides an infrared cut-off at this critical scale.

As a consequence of theorem 1, the density $\rho(\lambda,e)=\displaystyle \lim_{L\rightarrow\infty}\rho^{(L)}(\lambda,e),$ where 
\begin{equation}
\rho^{(L)}(\lambda,e)=L^{-d} \sum_{{\bf k},\sigma} \Xi^{(L)}(c^+_{{\bf k}\sigma}c_{{\bf k}\sigma};e,\lambda),
\end{equation}
is well-defined in the sense of formal power series. In the thermodynamic limit, the sum over quasi-momenta becomes on an integral over the first Brillouin zone,
\begin{equation}
\rho(\lambda,e)=\sum_\sigma \int \frac{d^dk}{(2\pi)^d}\eta_\sigma({\bf k};e,\lambda).\label{Definition_Density}
\label{eq: density}
\end{equation}
We prove the following statement:

{\bf Theorem 2:}
 The density (\ref{Definition_Density}) is independent of the interaction strength,
\begin{equation}
\rho(\lambda,e)=\rho(0,e)=2\vol{\Sigma_{e,\mu}},\label{eq: thm2}
\end{equation}
denoting by $\vol{\Sigma}$ the volume enclosed in the surface $\Sigma$.

\underline{Proof:} First observe that, because of the regularized dispersion relation, the ground state $\Omega^{(L)}$ of $H^{(L)}$ is non-degenerate, providing the coupling constant is small enough. It follows that the expectation value of an operator $\A$ is given by its ground state expectation, $\Xi^{(L)}( \A ,\lambda)=\langle\Omega^{(L)},\A\Omega^{(L)}\rangle$.

Since the number operator $N$ commutes with the Hamiltonian (\ref{HamiltonL}), $\Omega^{(L)}$ is an eigenstate of both $H^{(L)}$ and $N$. The eigenvalues of $N$ are positive integers, and the expectation value of the number operator is an integer.

On the other hand, $\Xi^{(L)}( N,\lambda)$ is analytic in $\lambda$ for small $\lambda$. Hence $\Xi^{(L)}( N,\lambda)$ is constant with respect to $\lambda$. In particular, $\rho^{(L)}(\lambda,e)=\Xi^{(L)}( N,\lambda)/L^d$ is independent of $\lambda$, and writing $\rho^{(L)}(\lambda,e)=\sum_{n\ge 0}\rho_n^{(L)}(e)\lambda^n$, we obtain $\rho_n^{(L)}(e)=0$ for $n\ge 1$. Taking the limit $L\rightarrow\infty$ ends the proof of the first equality in Eq.~(\ref{eq: thm2}). 

The second equality in Eq.~(\ref{eq: thm2}) is trivial: In a system of non-interacting fermions with a dispersion relation 
$e({\bf k},\mu)$, the occupation number is $1$, if $e({\bf k},\mu)\le 0$ and vanishes for $e({\bf k},\mu)>0$. 
The second equality in Eq.~(\ref{eq: thm2}) follows then from Eq.~(\ref{eq: density}) with $\lambda=0$.

We finally come to the proof of Luttinger's theorem.
We now follow the original argumentation of Luttinger in~\cite{L} to conclude that, if the density is fixed to its 
physical value $\rho_{phys}$, the volume enclosed in the Fermi surface is independent of the interaction strength.

In the free fermion approximation, obtained by setting $\lambda=0$, the (free) Fermi surface is given by
\begin{equation}
S_{0,\mu_0}=\{{\bf k}:\, \varepsilon({\bf k})=\mu_0\},
\end{equation}
where $\mu_0$ is chosen such that the density of the physical system satisfies
\begin{equation}
\rho_{phys}=2\vol{S_0}.\label{eq: Fixing chem pot 1}
\end{equation}
Consider now the system of interacting fermions.
The occupation number of the system of many interacting fermions, given by 
$n_\sigma({\bf k};\mu,\lambda)=\eta_\sigma({\bf k};E(\,\cdot\, ,\lambda,\mu),\lambda)$,
has a jump on the surface $\Sigma_{E(\,\cdot\, ,\lambda,\mu)}$.
$E(\,\cdot\, ,\lambda,\mu)$ is the interacting dispersion relation defined in~(\ref{EdispersionRelation}).

By definition the physical Fermi surface $S_{\lambda,\mu}$
is the surface on which the occupation number $n_\sigma({\bf k};\mu,\lambda)$ 
has a jump, and we can thus write 
\begin{equation}
S_{\lambda,\mu}=\Sigma_{E(\,\cdot\, ,\mu,\lambda)}.\label{eq: physical FS}
\end{equation}
The density of the interacting system is given by $\rho(\lambda',E(\,\cdot\, ,\lambda,\mu))$, 
for $\lambda'=\lambda$. By the theorem~2, 
with $\Sigma_{E(\,\cdot\, ,\mu,\lambda)}$ replaced by $S_{\lambda,\mu}$ and $e(\,\cdot\, ,\mu)$ replaced by $E(\,\cdot\, ,\lambda,\mu)$,
\begin{equation}
\rho(\lambda',E(\,\cdot\, ,\lambda,\mu))=\rho(0,E(\,\cdot\, ,\lambda,\mu))=2 \vol{S_{\lambda,\mu}}.
\label{eq: rho is vol S}
\end{equation}
In particular, the density of the interacting system is given by twice the volume enclosed in the 
physical Fermi surface.

As in the case of free fermions, the chemical potential $\mu$ has to be adjusted to its physical value
$\mu_{phys}=\mu(\lambda,\rho_{phys})$ in such a way that the density is fixed to its physical value,
\begin{equation}
\rho\left(\lambda,E(\,\cdot\, ,\lambda,\mu_{phys})\right)=\rho_{phys}.\label{eq: Fixing chem pot 2} 
\end{equation}
Equating the right hand side of Eq.~(\ref{eq: Fixing chem pot 1}) with the left hand side of 
Eq.~(\ref{eq: Fixing chem pot 2}), and using Eq.~(\ref{eq: rho is vol S}) leads to Luttinger's theorem,
\begin{equation}
\vol{S_{\lambda,\mu_{phys}}}=\vol{S_{0,\mu_0}}.\label{eq: Luttinger Thm}
\end{equation}
The interaction will in general deform the Fermi surface, and by the way change its volume. If the density
is kept constant, and since the density is always given by the volume enclosed into the Fermi surface according to
Eq.~(\ref{eq: rho is vol S}), the chemical potential has to be modified in order to compensate for the change in the 
volume of the Fermi surface.

As explained in the introduction, the present proof of Luttinger's theorem is perturbative, in the sense that only formal power series are considered. A non-perturbative proof of Luttinger's theorem, using the Fermi liquid construction of~\cite{FKT}, is probably possible, under the hypotheses of~\cite{FKT}. However, it turns out to be very difficult to prove the invertibility of the renormalization map non-perturbatively.

\acknowledgments

A. Praz thanks C. Mudry for stimulating discussions, and careful reading of our 
manuscript.

\end{document}